\documentclass[10pt, conference, compsocconf]{IEEEtran}
\usepackage{graphicx}
\usepackage{cite}
\usepackage{url}
\usepackage[cmex10]{amsmath}

\usepackage{algorithmic}
\usepackage{algorithm}
\usepackage{gensymb}
\let\subcaption\relax
\usepackage{caption}
\usepackage{subcaption}
\usepackage{graphicx}

\usepackage{stfloats}
\usepackage{footnote}
\ifCLASSINFOpdf
\usepackage[T1]{fontenc}
\usepackage[latin9]{inputenc}

\usepackage{multirow}
\PassOptionsToPackage{bookmarks=false}{hyperref}

\begin{document}

	%\parskip 0ex plus 0.2ex minus 0.1ex
	%\def\baselinestretch{0.99}
	
	\title{IBeaconMap: Automated Indoor Space Representation for Beacon-Based Wayfinding}
	
	\author{\IEEEauthorblockN{Seyed Ali Cheraghi, Vinod Namboodiri, Kaushik Sinha}
		\IEEEauthorblockA{Department of Electrical Engineering and Computer Science\\
			Wichita State University, Wichita, KS, USA\\
			Email: sxcheraghi@shockers.wichita.edu, \{vinod.namboodiri, kaushik.sinha\}@wichita.edu}}
	\maketitle
	
\begin{abstract}
	
Traditionally, there have been few options for navigational aids for the blind and visually impaired (BVI) in large indoor spaces.
Some recent indoor navigation systems allow users equipped with smartphones to interact with low cost Bluetooth-based beacons deployed strategically within the indoor space of interest to navigate their surroundings. A major challenge in deploying such beacon-based navigation systems is the need to employ a time and labor-expensive beacon planning process to identify potential beacon placement locations and arrive at a topological structure representing the indoor space. This work presents a technique called IBeaconMap for creating such topological structures to use with beacon-based navigation that only needs the floor plans of the indoor spaces of interest. IBeaconMap employs a combination of computer vision and machine learning techniques to arrive at the required set of beacon locations and a weighted connectivity graph (with directional orientations) for subsequent navigational needs. Evaluations show IBeaconMap to be both
fast and reasonably accurate, potentially proving to be an essential tool to be utilized before mass deployments of beacon-based indoor wayfinding systems of the future. 
\end{abstract}

%%%%%%\begin{CCSXML}
%%%%%%<ccs2012>
%%%%%% <concept>
%%%%%%  <concept_id>10010520.10010553.10010562</concept_id>
%%%%%%  <concept_desc>Computer systems organization~Embedded systems</concept_desc>
%%%%%%  <concept_significance>500</concept_significance>
%%%%%% </concept>
%%%%%% <concept>
%%%%%%  <concept_id>10010520.10010575.10010755</concept_id>
%%%%%%  <concept_desc>Computer systems organization~Redundancy</concept_desc>
%%%%%%  <concept_significance>300</concept_significance>
%%%%%% </concept>
%%%%%% <concept>
%%%%%%  <concept_id>10010520.10010553.10010554</concept_id>
%%%%%%  <concept_desc>Computer systems organization~Robotics</concept_desc>
%%%%%%  <concept_significance>100</concept_significance>
%%%%%% </concept>
%%%%%% <concept>
%%%%%%  <concept_id>10003033.10003083.10003095</concept_id>
%%%%%%  <concept_desc>Networks~Network reliability</concept_desc>
%%%%%%  <concept_significance>100</concept_significance>
%%%%%% </concept>
%%%%%%</ccs2012>  
%%%%%%\end{CCSXML}
%%%%%%
%%%%%%\ccsdesc[500]{Computer systems organization~Embedded systems}
%%%%%%\ccsdesc[300]{Computer systems organization~Redundancy}
%%%%%%\ccsdesc{Computer systems organization~Robotics}
%%%%%%\ccsdesc[100]{Networks~Network reliability}


%
%  Use this command to print the description
%
%\printccsdesc

% We no longer use \terms command

%%%%%%\keywords{ACM proceedings; \LaTeX; text tagging}

\section{Introduction}
Wayfinding can be defined as knowing where a person are in a building or an environment, knowing where their desired location is, and knowing how to get there from their present location. For outdoor environments, recent advances in global positioning systems (GPS) and mapping technologies provide accurate and simple to use means for wayfinding. For indoor environments, reading and following signs remains the easiest and most reliable option because GPS and associated advances for outdoor environments do not apply. This has, however, meant that indoor wayfinding has remained a challenge for the blind and visually impaired (BVI) in our society. Indoor environments can be geographically large and intimidating such as grocery stores, airports, sports stadiums, large office buildings, and hotels. A solution to the indoor wayfinding problem for the BVI also has broad applications for the sighted population. In unfamiliar, large indoor spaces, it is common for even typically sighted people to be disoriented and have trouble finding their way around. This could be due to the lack of well marked signs and maps, or not being familiar with the conventions or language used on these signage, or just the fact that the layout of the space is disorienting. 

%- Existing approaches and limitations

The most accurate and usable indoor wayfinding systems (e.g. \cite{Chang2010, Willis-Helal, Kulyukin2006}) available to persons with low vision used to be those that rely on the use of radio frequency identification (RFID) tag technology. This solution however is not very flexible when it comes to changing embedded information on tags; furthermore, the tag reader technology is expensive and can be difficult to integrate into current mobile systems. Other mechanisms that provide audible directions such as TalkingSigns \cite{talkingsigns} still need each user to possess special audio frequency devices capable of acting as receivers. In general, most approaches to solve this challenge require special hardware to be carried by the user. Such limitations have created barriers for widespread use and adoption for indoor wayfinding. Recent work has developed a system of wayfinding for the BVI using low-cost, stamp-size Bluetooth Low Energy (BLE) ``beacon'' devices embedded in the environment \cite{kim2016,GuideBeacon-percom2017,ahmetovic2016navcog2} that interact with smartphones carried by users. Such beacon-based navigation systems have achieved promising preliminary results indicating that they may be a viable solution for indoor wayfinding for the BVI if some of the underlying challenges to the deployment of such systems can be solved.  

%- indoor space representation need is acute, what is the input and output that we seek?%needs of beacon-based navigation, possible approaches and their challenges 
One major challenge facing beacon-based indoor wayfinding is that of creating fast and accurate representations of indoor spaces that can be used for beacon-placement and subsequent navigation computations\mbox{---}a process that we will refer to as \emph{beacon planning} in the rest of this paper. Manual determination of all beacon placement locations and path computations is time-consuming and labor-expensive, especially for large indoor spaces. Such an approach requires the manual identification of walking paths on a floor plan, marking of points of interest, determining the distance between any two points of interest, determining the orientation between them for navigation, computing shortest paths between points of interests, and subsequent adjustments to optimize the resulting paths that may require further iterations of the entire process. Other approaches of using robots or crowdsourcing are still time consuming and expensive/difficult to execute \cite{robotmapping,Chen-Crowdmap2015}.  

This paper presents the design of an indoor space representation technique for beacon planning called IBeaconMap that uses computer vision and machine learning techniques to automate the process of extracting the necessary information from an indoor space for subsequent beacon placement and path computations for navigation. IBeaconMap only needs as input an architectural floor plan of the indoor space of interest. It extracts all the points of interest (doors, stairs, elevators etc.), identifies walking paths within this space, and creates a connectivity graph representation of the space upon which path computations for navigation could be performed. In addition, IBeaconMap provides the exact 2-dimensional locations where beacons can be placed on the indoor floor plan. Evaluations of IBeaconMap show that it can create space representations that are reasonably accurate. Given that accuracies obtained vary based on floor plan quality and feature density, IBeaconMap can be configured to employ one or more feature detection options to identify the best fit. Best of all, it can typically provide its outputs within the order of a few minutes in most cases on common off-the-shelf computing devices as opposed to a time consuming and labor-expensive manual indoor space representation exercise. Even after beacons are placed through some mechanism (manual or some sort of an automated technique) there will always be a need to tag these beacons with the right information and make changes to beacon locations or information and re-compute the necessary data structures; IBeaconMap is built as an easy to use tool to achieve these post-initial deployment administration objectives as well.

% State of the Art in Indoor Navigation and why beacon-based navigation is best
\section{State of the Art}
This section surveys the various options available for indoor wayfinding and describes related work in the area of indoor space representations for wayfinding.
\subsection{Indoor Wayfinding Mechanisms}
Indoor wayfinding for the BVI does not require knowing the user's location at all times; rather it is more important to identify strategic points within an indoor space that a user should be localized at accurately (within 1-2 m localization error). The direction of using existing infrastructure in indoor spaces recently has largely revolved around using Wi-Fi access points (APs) that are already present. Under various assumptions, prior work has shown accuracies within a few meters (for e.g. \cite{Chronos2016, Kumar2014-Ubicarse, Xiong2013-ArrayTrack, Xiong2015-ToneTrack, Zaruba2007, Kotaru2015-SpotFi, Chintalapudi2010}). Although this direction achieves indoor wayfinding without any additional infrastructure costs, and allows users to use mobile devices they carry, they have many limitations in terms of AP density requirements, gestures required from users (which can be difficult for BVI users), and best accuracy possible. 

The direction of adding additional infrastructure in indoor spaces for localization has been explored in literature, primarily because of their promise of higher accuracies (compared to Wi-Fi based systems for example). Such work has included the use of technologies such as RFID (e.g. \cite{Wang2013-RFID}, Ultra Wideband (UWB) (e.g. \cite{Gezici05-UWB}), Ultrasound (e.g. \cite{Priyantha2000Cricket-Ultrasound}), Infrared (IR) (e.g. \cite{Aitenbichler2003-IR}), and visible light (e.g. \cite{Kuo2014-VLC}). Many of these technologies (some specific to indoor wayfinding for BVI such as \cite{talkingsigns, Chang2010}) are not effective for indoor wayfinding (and have rarely been used) because of the requirement of carrying additional hardware on the user, or more expensive or power-inefficient reference nodes in the environment. There have also been many attempts to use computer vision techniques to read and understand signs (as pointed out in \cite{manduchi-cacm}) to assist with wayfinding for the blind and visually impaired; these tend to have high inaccuracies in the information read out when a user is mobile and text is not directly facing the user.

Bluetooth-based indoor localization is not new (e.g. \cite{BluetoothClassic1}), but it only gained traction after the revision in 2010 and the introduction of BLE. The work in \cite{BLEvsWiFI} compares BLE-based localization to WiFi-localization and show the the former is far more accurate than the latter. Other work with the Apple iBeacon platform showed accuracies as small as 0.53 m \cite{iBeaconPrimer} whereas others focused more on the techniques that can be used to improve localization accuracies \cite{BLELocalization2016}. There have been three recent works using BLE beacons for wayfinding for the BVI, StaNavi \cite{kim2016}, GuideBeacon \cite{GuideBeacon-percom2017}, and NavCog \cite{ahmetovic2016navcog2}. All of them report significant improvement in the ability of the BVI to navigate unfamiliar indoor spaces independently using the developed systems. Other efforts include the \emph{Wayfindr} project \cite{wayfindr} which is an effort to develop an open standard for navigation for the visually impaired in outdoor and indoor spaces, including the use of BLE beacons. Other systems for indoor navigation for the BVI using beacons such as \cite{indoors, blindsquare} are more for providing location updates as a user walks through, and are not turn-by-turn navigation systems such as GuideBeacon, NavCog and StaNavi.

\subsection{Indoor Space Representation}
Although there are currently various BLE beacon-based navigation systems being developed, none of the efforts so far had designed a fast, and largely automated method for representing indoor spaces as topological structures for accurate and timely beacon-based navigation. Such a method (as proposed in this work) will benefit all current efforts in deploying indoor beacon-based navigation systems. Work related to creating representations of indoor spaces have been around for a while (e.g. \cite{CityGLM,Kruminaite:2014,Niua2016,Whiting2007,Becker2009,Setalphruk2002}). These can mainly be differentiated based on the approach used in collecting the required information and in the techniques employed to create the desired representations. IBeaconMap differs from this class of work by taking files in simple image formats or PDFs as input and employing a combination of computer vision and machine learning techniques. In addition to marking points of interests on floor plans as beacon locations, IBeaconMap can also mark strategic points such as intersections which are important for BVI navigation. None of the previous work on indoor space representations focused on providing outputs catering to the special needs for beacon-based wayfinding that include beacon location markings, indoor paths connecting these locations, a weighted connectivity graph as topological structure representation, and directional orientations for paths. The web-based mapping tool developed as part of NavCog \cite{ahmetovic2016navcog2}, the only other tool with a similar objective as IBeaconMap, requires a user to mark all beacon locations and walking paths first on a floor plan image. This higher-level of manual involvement is expected to not scale well thus rendering the tool not as desirable in many situations.

Additional approaches for indoor space representations beyond extraction from architectural floor plans are that of robotic mapping and crowdsourced approaches \cite{robotmapping, puikkonen-indoormapcasestudy, Chen-Crowdmap2015, Rai-Zee-2012}. Robotic mapping approaches are likely to be more expensive to implement and time-intensive while crowdsourced approaches, although inexpensive and maybe even free, will not be as accurate or fast as IBeaconMap. Further, the recent work on using crowdsourcing to deploy beacons in \cite{gleason2017luzdeploy} assumes that beacon locations are already known; thus, IBeaconMap could be a useful first tool to create location markings where beacons can then be placed in a crowdsourced fashion.

% The GuideBeacon System 
%- Overarching details, 
%- Evaluation methodology and results
\section{The Beacon Planning Challenge}
This section provides a more detailed description of the beacon planning challenge for beacon-based wayfinding. The workings of a typical beacon-based wayfinding system is described first followed by some possible beacon planning approaches and the challenges faced therein, thus motivating the proposed IBeaconMap tool. Note that the term ``beacon planning'' here refers to the entire process of characterizing indoor spaces, determining points of interest (PoIs), creating connectivity graphs and performing any pruning operations, finally arriving at locations at which beacons should be placed. It does not refer to the actual physical placement of the beacons (called beacon placement), subsequent testing and optimizations of received signal strengths from the BLE devices and configuration of parameters like messaging intervals and transmission power levels, all of which may be required.

\subsection{Working principle of beacon-based wayfinding systems}

\begin{figure}[htbp]
	{
		\includegraphics[width=3.4in,keepaspectratio]{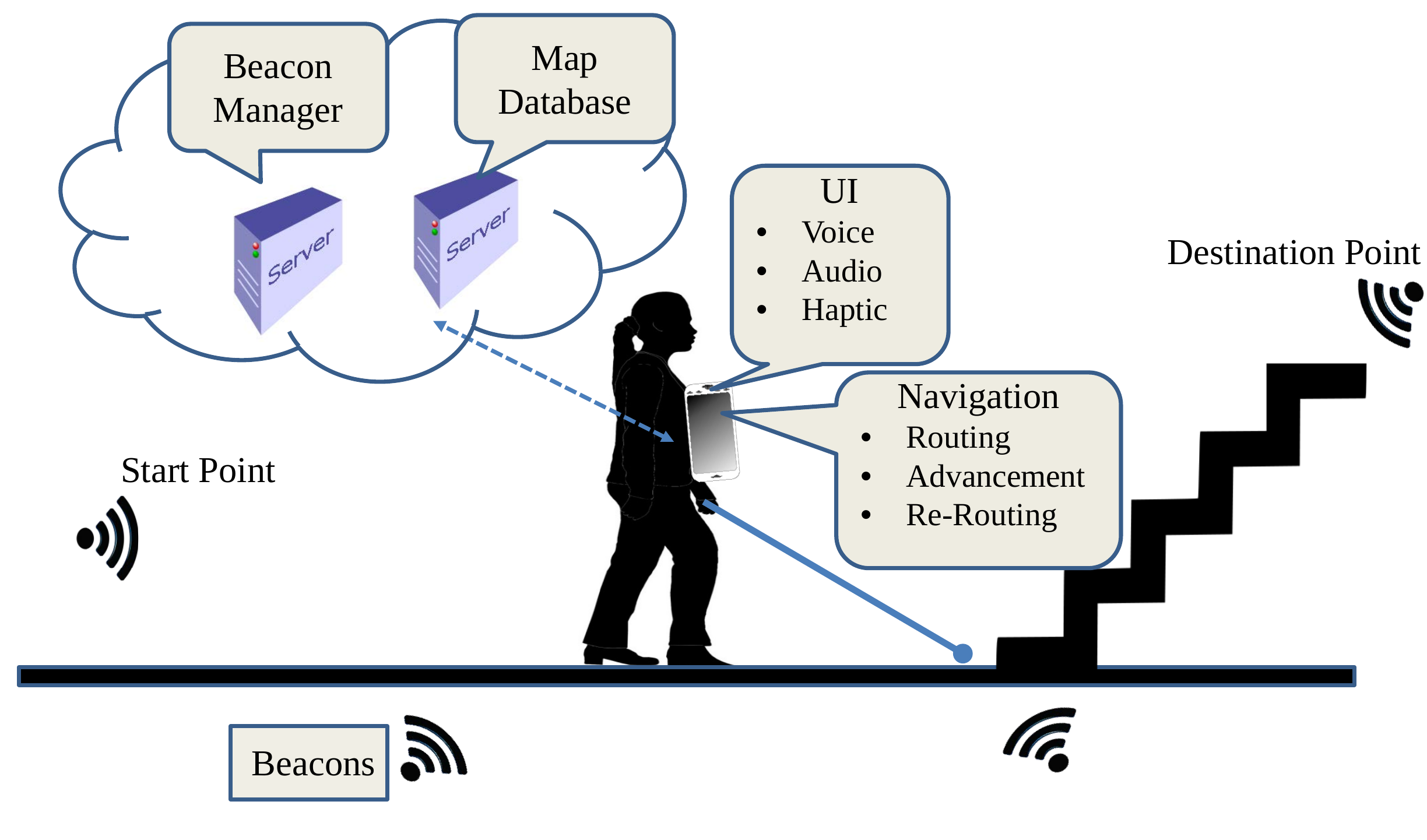}
		\caption{Building blocks and interactions of the GuideBeacon system \cite{GuideBeacon-percom2017}.}
		\label{fig:SystemDiagram}
	}
\end{figure}

Upon entering an indoor space and starting the wayfinding application on their smartphones, users are prompted to provide their desired destination (typically using voice commands for the BVI). The specified destination is then looked up in a database of points of interest (PoIs) in the indoor space (provided typically by the beacon manufacturer's platform as a Beacon Manager on a server); if matches are found, they are listed out to the user one by one until the user confirms one of them. Upon confirmation that there is a match for the desired destination, the system then downloads information about the connectivity graph of the indoor space on which path computations are performed to arrive at an end-to-end path the user utilizes to navigate to the destination. Along the route, the smartphone detects proximity to the beacons to determine the user location and provide the next steps they must take to stay on the route. Additional optimizations for advancement along the route are possible to better improve the detection time and accuracy. The overarching components involved in the GuideBeacon system are reproduced in Figure \ref{fig:SystemDiagram} so that the reader gets a better sense of where beacon planning fits in any beacon-based wayfinding system. 

BLE beacons are typically placed strategically in the indoor space and include both PoIs and other points that may be useful to improve the navigation experience. Thus, in a connectivity graph of the indoor space, PoIs will only be a subset of all beacons. All beacons could serve as vertices of this graph, and paths between them as edges. To compute useful end-to-end routes, the physical distance and difficulty level of a path must be encoded as a weight to the connectivity graph; a shortest-path algorithm could then return the most favorable path for a user. In addition to characterizing path difficulty as weights, the directional orientation of paths must be stored so that users can be directed to their next point along the route to the destination.

\begin{figure*}%[!tbp]
	\centering
	\begin{minipage}{0.33\textwidth}
		\includegraphics[height = 2.2in, width=\textwidth]{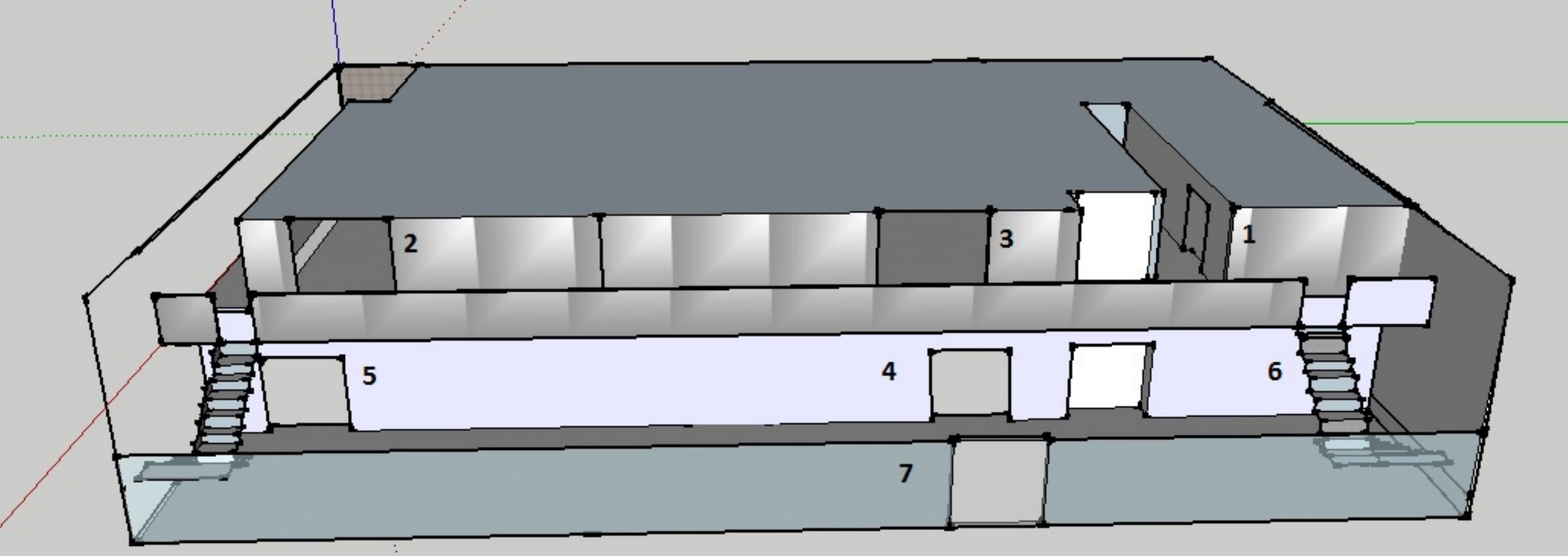}\label{fig:f2}
		\subcaption{Front view (facing North)}
	\end{minipage}
	\begin{minipage}{0.33\textwidth}
		\includegraphics[height = 2.2in, width=\textwidth]{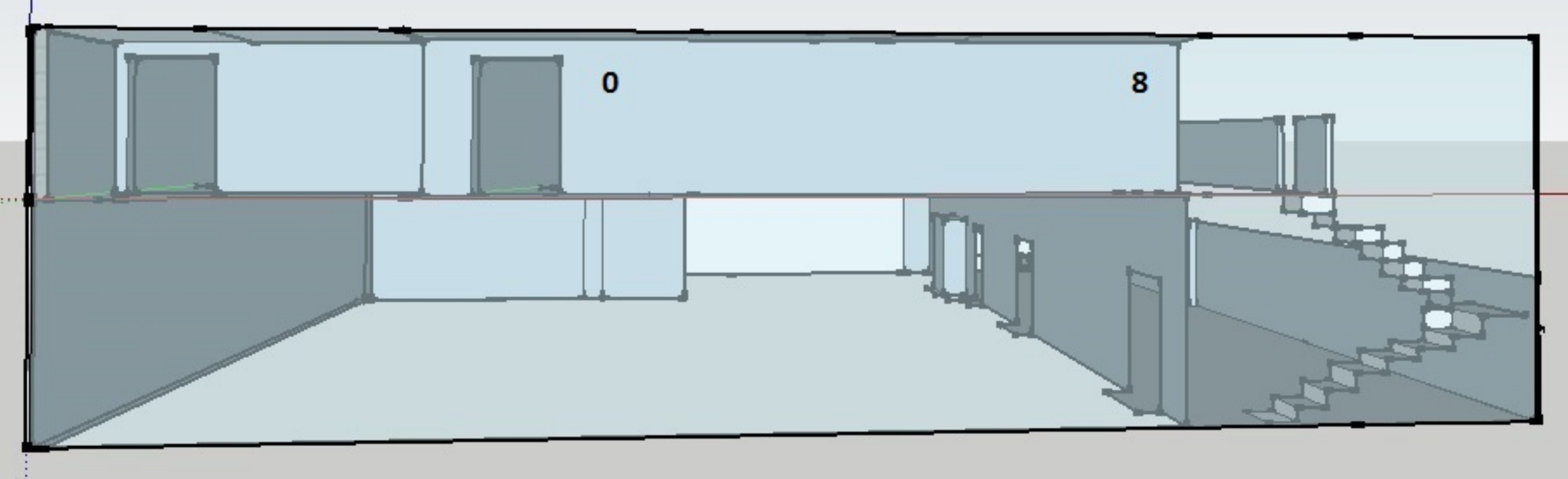}\label{fig:f1}
		\subcaption{View from the left (West) facing East}
	\end{minipage}
	\begin{minipage}{0.25\textwidth}
	\includegraphics[height = 1.75in, width=\linewidth]{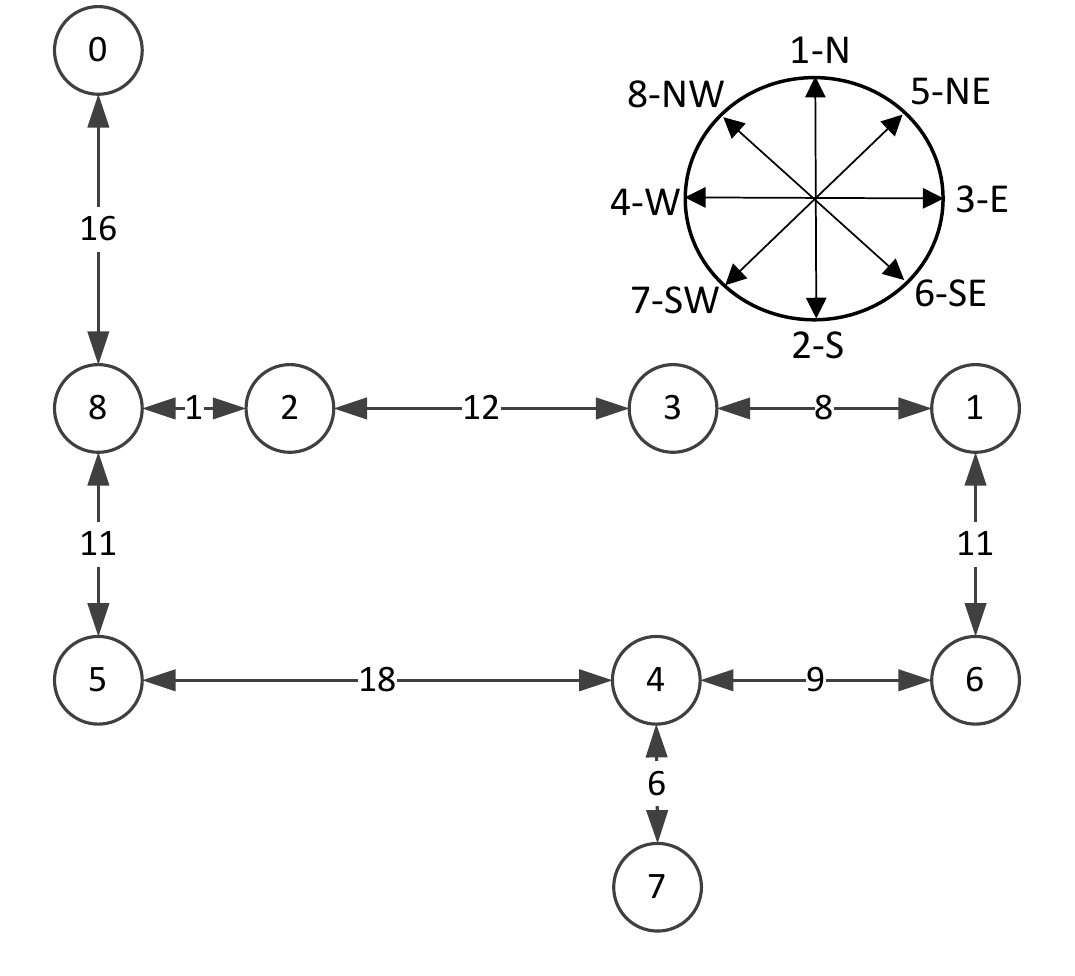}\label{fig:graphcompass}
	\subcaption{Beacon deployment within space modeled as a weighted graph; also shows a numeric code assigned to various possible compass orientations.}
		\end{minipage}
	\caption{Beacon deployments inside a building and resulting weighted connectivity graph and orientations along paths}
	\label{fig:SpaceandGraph}
\end{figure*}
%
%\begin{figure}[!ht]
%	\centering{
%		\includegraphics[height = 2.5in, width=0.75\linewidth]{networkMap.pdf}}
%	\caption{Building beacon deployment of Figure \ref{fig:3Dpictures} modeled as a graph showing vertices, edges, and weights. The top right also shows a numeric code assigned to various possible compass orientations.}
%	\label{fig:graphcompass}
%\end{figure}

\subsection{Manual beacon placement} \label{subsec:manual}

Any indoor space can be prepared for beacon placement manually (as in GuideBeacon\footnote{The mapping tool provided with NavCog \cite{ahmetovic2016navcog2} is similarly a manual process requiring a human to work with a software tool to mark out all beacon locations first and then walking paths before connectivity information can be generated. StaNavi \cite{kim2016} did not specify beacon planning or placement in their work.}) as follows. All possible PoIs are marked as beacon locations. On a floor plan image or sketch, lines are drawn through the middle of all walking paths (including stairs) in the indoor space on a 2D basis. All intersection points between lines are marked as beacon locations. On each line where marked beacon locations are more than $x$ feet apart, mark additional beacon locations every $\frac{y}{\lceil \frac{y}{x}\rceil}$ from one of the PoIs, where $y$ is the current separation of PoIs in feet before this step. At the end of these steps, the indoor space has all beacon locations marked out at which point beacons can be placed at a location most convenient near to it. This, however, is not the end of the process. Using the beacon locations marked out, a connectivity graph has to be subsequently created with measured distances between vertices (beacon locations) possibly serving as weights between the edges.\footnote{Other weight measures can be expected foot traffic, difficulty of terrain which we do not explore at this point} and orientations. Additional optimizations (such as pruning of the graph) may require an iterative approach to arrive at the final topological structure to use for navigation.

Figure \ref{fig:SpaceandGraph}a and b show a 3D representation of an indoor space from two different angles. It also shows the assignment of 9 beacons numbered from 0 through 8 so that a person can navigate from the building entrance to a room on the left side of the second floor. Figure \ref{fig:SpaceandGraph}c illustrates a conversion of the 3D space and its beacon to a directed graph with beacons serving as vertices and paths serving as edges. The edge weights are the distances on the paths between beacons. The figure also shows a numeric code representing orientation; each possible direction is placed into one of either different bins of size 45$\degree$ centered at the positions shown. 

Determining beacon locations in an indoor space including obtaining weights and orientations to use on paths is thus currently a time and labor-intensive process. As the scale of the indoor space increases (for example a floor plan of a large indoor facility as shown in Figure \ref{fig:metroplex} later), the number of PoIs in the space typically tends to increase at least linearly, rendering a manual beacon planning process as a big challenge. To solve this challenge, this work proposes the IBeaconMap technique next that can (largely) automate the beacon planning process needing only an architectural floor plan image as input. 

% Methods
\section{Beacon Planning with IBeaconMap} 
The IBeaconMap technique involves taking a two dimensional floor plan as input and providing a connectivity graph as output with all the necessary information (beacon locations, weights, direction/orientations) required for subsequent navigation. In this work we only discuss how one given floor plan of an indoor space can be represented; there may be multiple such floor plans that will need to be analyzed (and merged) to arrive at the final representation for an entire building.\footnote{In architectural drawings, indoor spaces are represented as individual floors with connection points shown on each floor plan.} This section begins by describing the input requirements to IBeaconMap followed by the presentation of the details of how IBeaconMap arrives at the desired output. 

\subsection{Input Requirements}\label{subsec:input}
There are two types of inputs required for IBeaconMap; one the floor plan itself, and second the set of building blocks (comprised of a set of features) that the algorithm is expecting to find on the image. 
\subsubsection{Floor Plan}
A floor plan is a drawing to scale to show a indoor space's details such as rooms, stairs etc. from a top view. It can be in various file extensions (PDF, JPEG, BMP, etc. for images and .DXF as a data format) and drawn using popular tools such as AutoCAD Architecture \cite{autocad-arch}. A typical floor plan could show interior walls and hallways, restrooms, windows and doors, interior amenities such as fireplaces, saunas and whirlpools, rest areas, service counters and many other PoIs within buildings. It also shows an indoor space's orientation and to what scale the floor plan is drawn. Figure  \ref{fig:fp1original} shows an example of a floor plan. With a scale factor, it is possible to extract distances between any two PoIs from the floor plan image by factoring in image resolution and the number of pixels that separate them. When the term ``high resolution'' is used in this work, it refers to those that are 2200 by 3400 pixel or better, common for non-scanned images. As will be seen later, IBeaconMap uses machine learning techniques to offset the weaknesses of image processing techniques at lower image resolutions. Because IBeaconMap can work well on low resolution images, it can also be easily applied to scanned images of hand-drawn sketches of architectural floor plans. This makes IBeaconMap applicable to floor plans that do not come from modern architectural drawing tools.  

\subsubsection{Building Blocks}
Building blocks are a set of features shown by a specific symbol and are important elements to be identified for subsequent indoor wayfinding as they tend to be PoIs where beacons are placed. These symbols are usually common between various architectural drawings and therefore a database of building blocks can be constructed and harnessed by an application trying to locate them on a floor plan. Building blocks can be categorized into various groups which can be prioritized based on the application needs. For example, those building blocks which are used to represent features related to indoor walking paths can be in the first group and summoned first. To be computationally efficient (reduced processing time), building blocks to check for on a floor plan can be selected based on the type of space under consideration; typically a pool or sauna is not expected to be seen in a research building, but doors, stairs, elevators, and other building blocks are expected.   

\subsection{Image Analysis}
A floor plan image's analysis is performed in four phases. In phase 1, the indoor path within the space is identified and removed for further analysis. In phase 2, all building blocks (PoI candidates) within the remaining floor plan image are identified. In phase 3, a skeleton of the indoor path is generated with adjoining building blocks mapped onto it. Finally, in phase 4 a connectivity graph is created through a traversal of the skeleton. 

\begin{algorithm}[ht]
	\caption{IBeaconMap Algorithm}
	\label{alg:indoornavimap}
	\begin{algorithmic}[1]
		\STATE \textbf{Input}: Floor plan image, set of building blocks, scale factor, and map orientation. Restricted areas on floor plan are marked before in pre-processing steps. Option of indoor walking path only or full floor plan detection is specified.
		\STATE \textbf{Result}: Connectivity graph with distance-based weights and compass orientation of each edge
		\STATE $R_{i} = regionprops~(FloorPlan,`Area')$ \{Using `Area' as a property to label the actual number of connected pixels in each region of the floor plan and save them in an array\}
		\STATE $R_{indoorpath}=Max(R_{i})$ \{Choose the region with maximum area\}
		\STATE $Flr_{f} = detectfeatures(FloorPlan)$ \{Return an array with various features of the input floor plan\}
		\STATE $(x,y)_{j} = classifyfeatures(Flr_{f},option)$ \{Classify features using specified technique as option (1: FDM, 2:FDM+SML, 3:FD + SML) in floor plan as building blocks of interest and return pixel indices\}
		\STATE $(x,y)_{indoorpath} = sel((x,y)_{j},R_{indoorpath})$ \{Select matched building blocks which are in the indoor path\}
		\STATE $skel = bwmorph(R_{indoorpath}, `skel')$ \{Morphological operation using `skel' as a property to remove pixels from the boundaries of the indoor path without breaking it apart\}
		\STATE $(x,y)_{skeleton} = Map((x, y)_{indoorpath},skel)$ \{Map $(x,y)_{indoorpath}$ on the indoor path skeleton\}
		\STATE $Path = BFS((x,y)_{skeleton})$ \{Determine connectivity information (distance in pixels, orientation) between every two adjacent nodes on the skeleton using a breadth first traversal, storing a string of orientation directions from each path\}
		\STATE Use image scale factor and resolution to find physical distances from any two building blocks to use as weights of the connectivity graph
	\end{algorithmic}
\end{algorithm}

\subsubsection{Phase 1: Indoor path identification}\label{subsubsec:phase1}
The goal of phase 1 (lines 3-4 in Algorithm \ref{alg:indoornavimap}) is to extract the indoor path and adjoining PoIs from the floor plan. The walking path connects all the building blocks (doors, stairs etc.) to each other, so finding it first makes it easier to find PoIs. Furthermore, having the indoor path helps find the shortest path from any office or point of interest to any other. Walking paths were found by identifying the largest contiguous block of pixels within the indoor space; this contiguous block of pixels has to be the walking path with all other areas within the floor plan having disconnections due to doors, walls, stairs etc. The largest contiguous area\footnote{Any area to be removed from consideration as a restricted area can be marked as such in floor plan pre-processing steps as described later in Section \ref{sec:tool}.} is then labeled so that it can be marked off as the walking path. Figure \ref{fig:fp1indoor} demonstrates the original floor plan  with the gray area added manually for illustration purposes to show the indoor path in the original image. In some scenarios a door or any other connected building block can be inside the area of the indoor path; these building blocks are not as important as PoIs on the walking paths for beacon-based wayfinding, but can still be found similarly. A user is provided an option (described later in Section \ref{sec:tool}) to choose between one of two beacon planning options: (i) find PoIs around indoor walking path only, and (ii) find PoIs from the entire floor plan. The first case is used when beacons are expected to be deployed only on areas surrounding the indoor walking path. This is expected to be the most common case where a person is guided all along and at the edges of indoor walking path, and once they enter through a door, further navigation is not required. The case of entire floor plan detection is more complex (and a worst case for IBeaconMap) as all building blocks on a floor plan need to be detected regardless of the proximity to the indoor path. This case arises when there may be multiple entryways within a walled area.  

\subsubsection{Phase 2: Building block detection}

Through the previous phase, a user selects the appropriate technique which leads to choosing part of the floor plan of interest (called foreground). After obtaining the foreground, the next step is to find all the required building blocks in a floor plan and get the specific coordinates of their locations (lines 5-8 in Algorithm \ref{alg:indoornavimap}). To achieve this goal, IBeaconMap provides three different approaches: feature detection and matching (FDM), feature detection, matching, and supervised machine learning (FDM + SML), and feature detection and supervised machine learning (FD + SML). The reason to use three different techniques is to provide options to users when faced with varying quality and complexities of floor plans supplied as input. The FDM approach is the fastest of the three, and is very accurate if the provided floor plans are of high resolution and without a high density of features. If the provided floor plan does not meet this criteria, as is possible when using scanned images of floor plan drawings made many decades ago, the accuracy can suffer. Having the other two approaches besides FDM provides more opportunities to arrive at an acceptably accurate result. The addition of SML to FDM allows removing some false positives from the FDM approach output, helping improve accuracy. For cases where FDM is expected to have very high inaccuracies, it can be skipped altogether. Instead, a pre-processing step of FD can be executed to first collect all possible features in the floor plan (a computationally intensive step) followed by SML to classify building blocks with reasonable accuracy. Each of these approaches are described next. 

\medskip
\noindent
\textbf{Feature Detection and Matching Approach}\\
Using object recognition and matching, common image processing techniques to detect features of an image, building blocks in the floor plan are  found. This process resembles image registration procedures which try to overlay two images from the same scene but from different angle or different sensors over each other. Two steps that are common are Image Feature Detection and Feature Matching. Feature detection is an image processing technique that has widely been used in the computer vision community. Image classification and retrieval, object recognition and matching, and texture classification are some of the areas that typically use feature detection. In this step salient and distinctive components of an object or an image such as corners, edges and etc. are detected. Feature or image matching is the process of matching detected features of the same scene between two images. It is part of many computer vision applications such as image registration and object recognition.
There are two approaches~\cite{features} to better understand the features in an image: area-based methods and feature-based methods. In the case of indoor buildings there are enough distinctive and detectable objects to be able to use the feature-based method from Shi and Tomasi ~\cite{shitomasi} that provides relatively good accuracy and low processing time. After detecting and extracting features, matching is performed next. By using an exhaustive method of determining the pairwise distance between features found from the floor plan, they can be matched over each other. The sum of squared difference (SSD) is used to measure the distance between features to perform this matching.

\medskip
\noindent
\textbf{Supervised Machine Learning after Feature Detection and Matching}\\
For cases where image resolution isn't high enough (defined earlier in Section \ref{subsec:input}), FDM may inaccurately classify certain features within floor plans as doors. In such cases the addition of supervised machine learning techniques can help eliminate some false positives. In this hybrid approach, after finding building blocks using FDM, those locations are cropped from the floor plan image. These cropped images of the floor plan with detected coordinates in their centers are given to a Support Vector Machine (SVM) \cite{SVM} (one approach to SML) for classification. As an efficient classifier, SVM uses given labeled training data to define an optimal hyperplane for classification purposes. This estimated hyperplane is used to classify the cropped images to one building block (doors in this case) versus all other building blocks found.

\medskip
\noindent
\textbf{Supervised Machine Learning after Feature Detection}\\
In cases where the matching process in FDM struggles badly to detect locations as designated building blocks, just feature detection (FD) can be used as a pre-processing step to collect all features from the supplied floor plan. Executing SML mechanisms on this set of collected features can improve accuracy of classification. One limitation with this approach is that the number of detected features on some floor plans can be very large (hundreds of thousands), making it computationally unappealing to run SML on. Thus, IBeaconMap employs a K-Means clustering technique \cite{kmeans} on densely populated features in a floor plan to reduce the candidate set of features to run SML on.  K-Means clustering technique is an unsupervised learning algorithm to find groups of unlabeled data. Since the exact number of clusters (PoIs) cannot be foreseen, the algorithm starts with a default number of clusters in a segment (half of detected features) which is then revised based on relative separation between clusters. Eventually, using SML, building blocks can then be sifted out from those symbols that are not building blocks. The three-step FD + SML approach takes more processing time than the other schemes, but can be more accurate than the others, especially with low resolution images and high feature density floor plans. 

 After obtaining building blocks' locations in terms of $(x,y)$ pixels using one of the above techniques, the next step is to separate and extract smaller areas from the floor plan (Image Segmentation). To achieve this goal a technique called image dilation is used where all the lines (walls, doors, etc) are made thicker; therefore, if there is any disconnection on a line, it is filled. Using the labeling method, removing areas which are smaller than a threshold, and removing the already located indoor path and the margin area, all the rooms can be extracted.
\subsubsection{Phase 3: Skeleton generation}
\begin{figure*}[ht]
	\centering
	\begin{minipage}{0.23\textwidth}
		\includegraphics[width=0.68\textwidth,angle = 90]{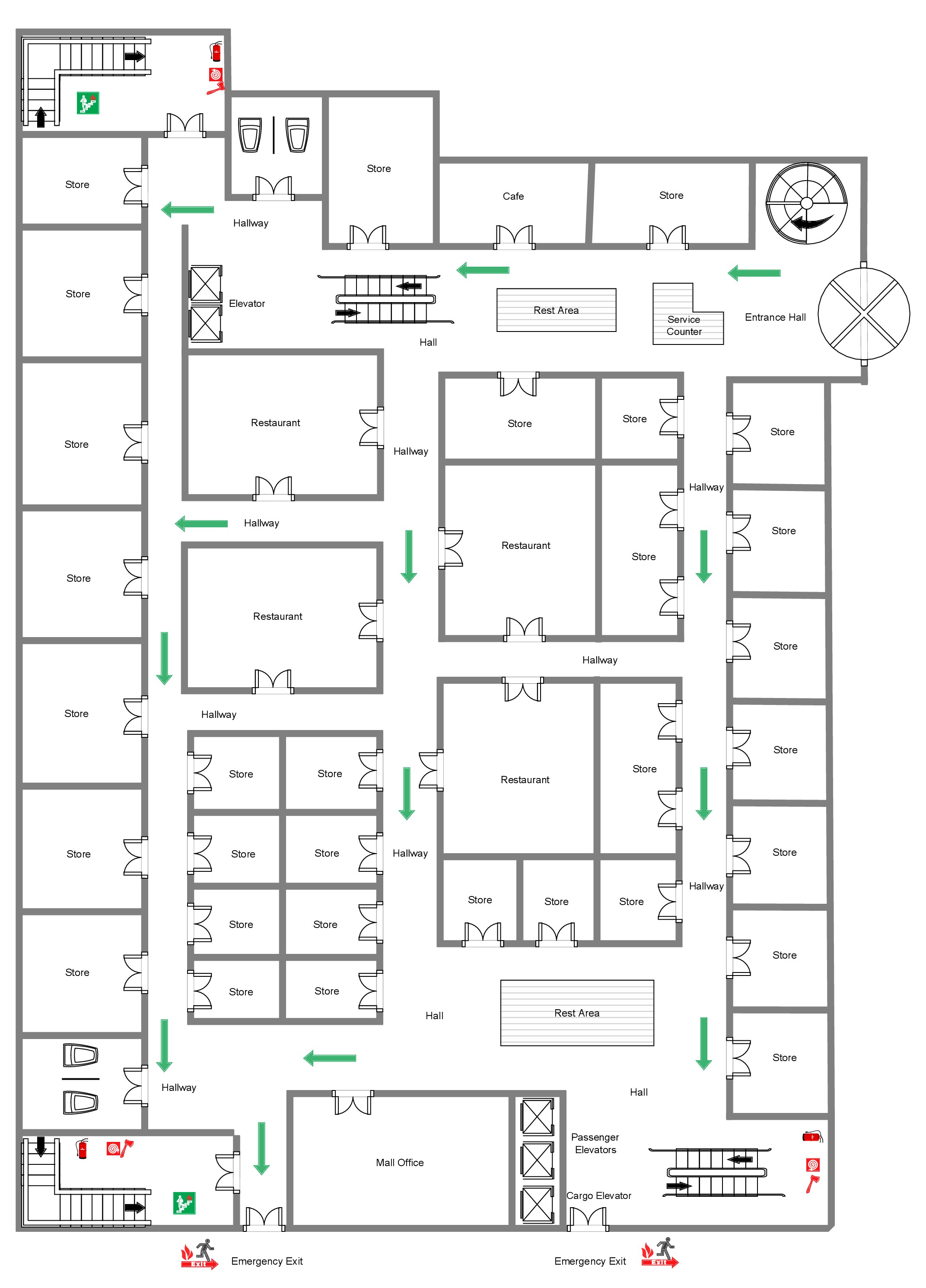}
		\subcaption{Original floor plan.}
		\label{fig:fp1original}
	\end{minipage} \quad
	\begin{minipage}{0.23\textwidth}
		\includegraphics[width=0.68\textwidth,angle = 90]{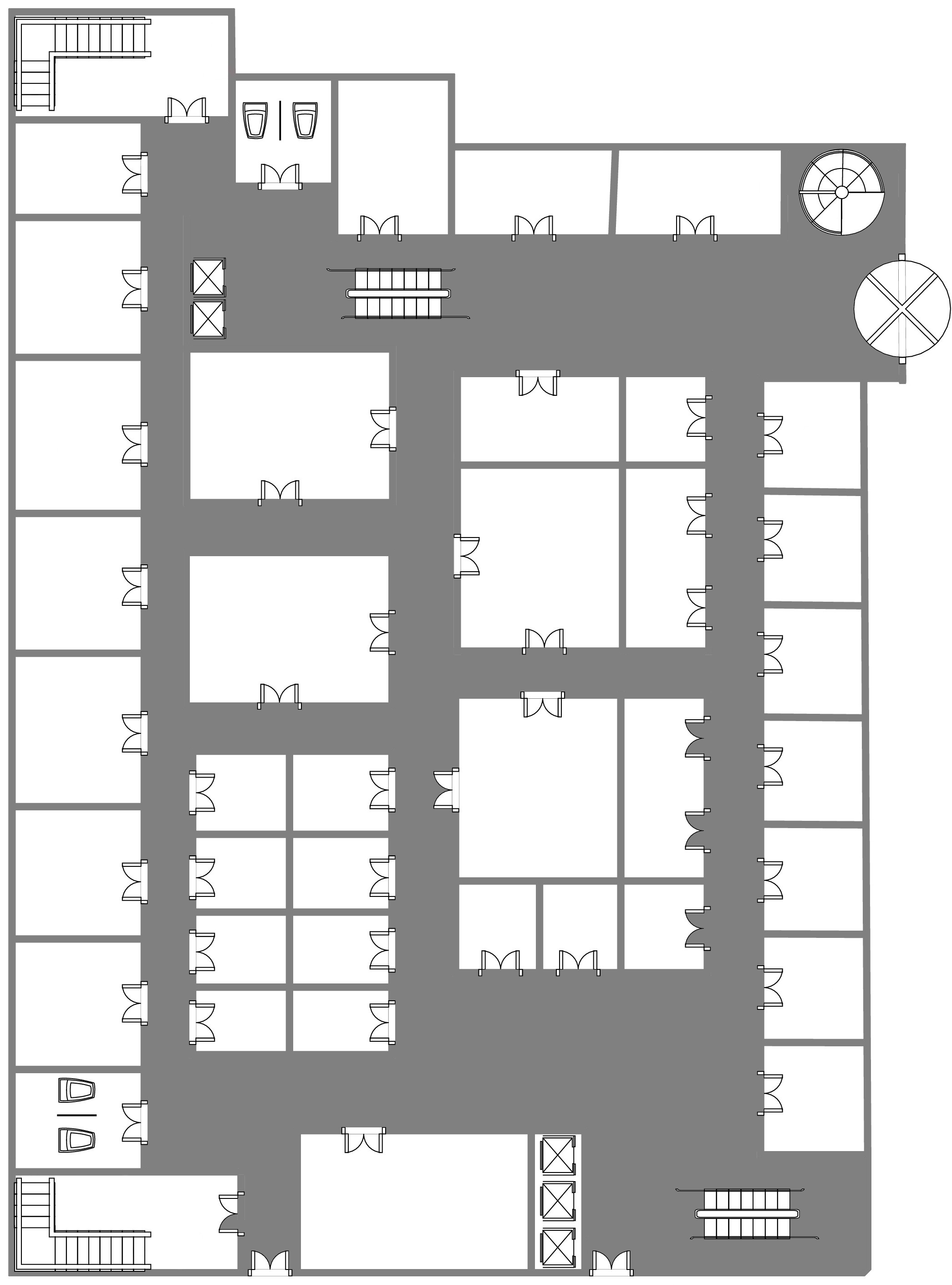}
		\subcaption{Extracted indoorpath.}
		\label{fig:fp1indoor}
	\end{minipage}
	\begin{minipage}{0.23\textwidth}
		\includegraphics[width=0.68\textwidth,angle = 90]{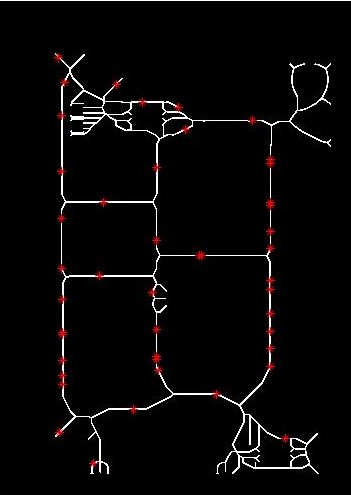}
		\subcaption{Skeleton output.}
		\label{fig:fp1skel}
	\end{minipage}\quad
\begin{minipage}{0.23\textwidth}
	\includegraphics[width=0.68\textwidth, angle = 90]{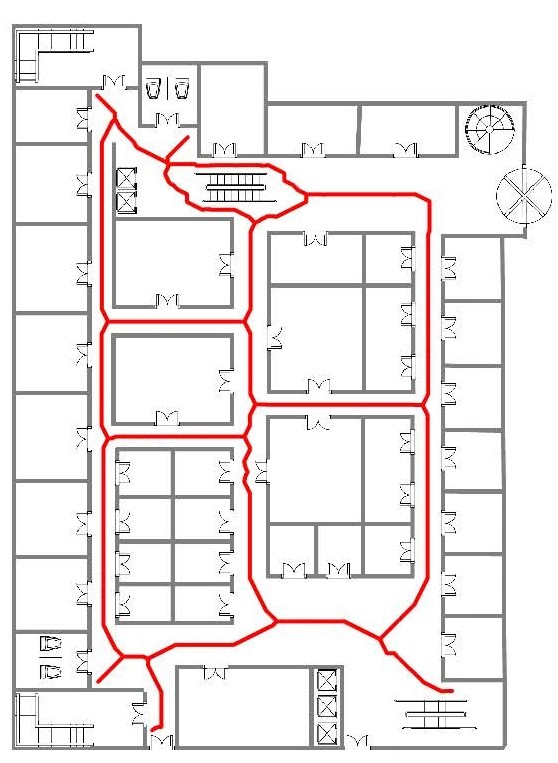}
	\subcaption{Final connectivity graph.}
	\label{fig:fp1graph}
\end{minipage} \quad	
	\caption{Outputs from execution of IBeaconMap on a floor plan.}
	\label{fig:fp1}
\end{figure*}

To connect one PoI to another, a path is required that does not pass through a wall, stair or any point which has a color other than white (after the floor plan is converted to a binary image). Since the locations of identified building blocks can be on the black line or be blocked in some ways, we desire to map them onto specific pixels of the indoor path already found. To achieve this, the boundary pixels of the indoor path are removed without letting the indoor path break apart. Then by using euclidean distance, the closest points on the indoor path skeleton to the building blocks are located (Figure  \ref{fig:fp1skel}). These steps are on lines 9-10 in Algorithm \ref{alg:indoornavimap}.\footnote{The function bwmorph used comes from the equivalent MATLAB function that was used to perform the skeleton generation operation.}

\subsubsection{Phase 4: Connectivity graph generation}
After mapping building blocks\footnote{If there were errors in building block detection, post-processing steps described later in Section \ref{sec:tool} can correct them; after such steps the connectivity graph generation phase is executed again.} on the indoor path skeleton, we need to find the paths connecting any PoIs which will lead to creating a connectivity graph on which path computations for navigation can be performed (lines 11-12 in Algorithm \ref{alg:indoornavimap}). For each path, the orientation required to move along the paths is also extracted and stored. Conversion from the pixels on the floor plan to actual distances requires scaling with the scale factor of the specific floor plan under consideration.

When a floor plan is taken by the IBeaconMap algorithm, the direction it faces is also provided as input. The floor plan image also provides the scale factor as input. For example, a scale of 1/16" = 1'- 0" results in each 1/16" (inch) on the plan as equivalent to one feet and zero inch of actual physical length. It is required that the image dimension (width and height) and both the horizontal and vertical resolution (for example 2200 * 3400, 200 dpi, 200 dpi) is provided as input for the algorithm.\footnote{200 dpi is equivalent to 200 pixels per inch; thus, every 200 pixels is 16 feet of physical distance.}

To determine one-hop path distances between PoI's, the IBeaconMap algorithm considers the indoor path skeleton to be the only non-zero pixels in the floor plan image. This by itself does not provide the one-hop paths between PoIs, but the skeleton can be traversed in a breadth-first fashion beginning from a PoI pixel by pixel to find various features. Any saved connectivity information includes the starting location, the path and the destination and the number of each of the characters involved (`E', `W', `N', `S', `NW', `NE', `SW', `SE') to arrive at orientation information for the path. The connectivity graph arrived at for the example floor plan under consideration is shown in Figure ~\ref{fig:fp1graph}.

%TODOs
%Where to describe security level, tool picture
% Is there an option to use on IP or hybrid approach?

% Evaluation Story

%Image processing, preferable approach
%- accuracy of beacon placement points
%-computation speed

% Limitations of using Image Processing Approach alone
%show example, accuracy numbers

% Hybrid approach used in IBeaconMap
% Show improved accuracy for case where image processing did not do well alone
% Discuss training methodology (perhaps beginning of evaluation?), overall time using both approaches

\section{Evaluation}
This section presents the results of evaluations of the effectiveness of IBeaconMap beginning with a description of the metrics used.  
\subsection{Metrics}
Two metrics were chosen to show the effectiveness of IBeaconMap. The first metric is that of accuracy of IBeaconMap's output in terms of the number of beacon locations correctly identified versus those that were incorrect. The incorrect ones are further broken down into beacon locations that were missed and those that were redundantly added. A correct identification of a beacon location involves finding a PoI and intersections. A visual comparison of beacon location marking outputs from a manual beacon planning process (as described earlier in \ref{subsec:manual}) is also presented to provide a visual sense of accuracy of IBeaconMap.

%in comparison to a similar output generated manually as used in \cite{GuideBeacon-percom2017}. If the beacon locations provided by IndoorBeaconMap are close enough to the manual outcome, it can be expected that the navigation experience would be similar as well. 

The second metric is the processing time for IBeaconMap to take a floor plan as input and produce its output. This metric is thus a measure of the reduction in time and labor in arriving at beacon locations and connectivity graph for indoor navigation. Any manual post-processing required to fix inaccuracies would need to be added onto this time for a fairer comparison with a completely manual process; however, the aim with IBeaconMap was to keep the manual corrections to be minimal.

\begin{figure*}
	%\centering
	\begin{minipage}{0.49\textwidth}
		\centering
		\includegraphics[width=0.5\textwidth, height = 3in, angle = 90]{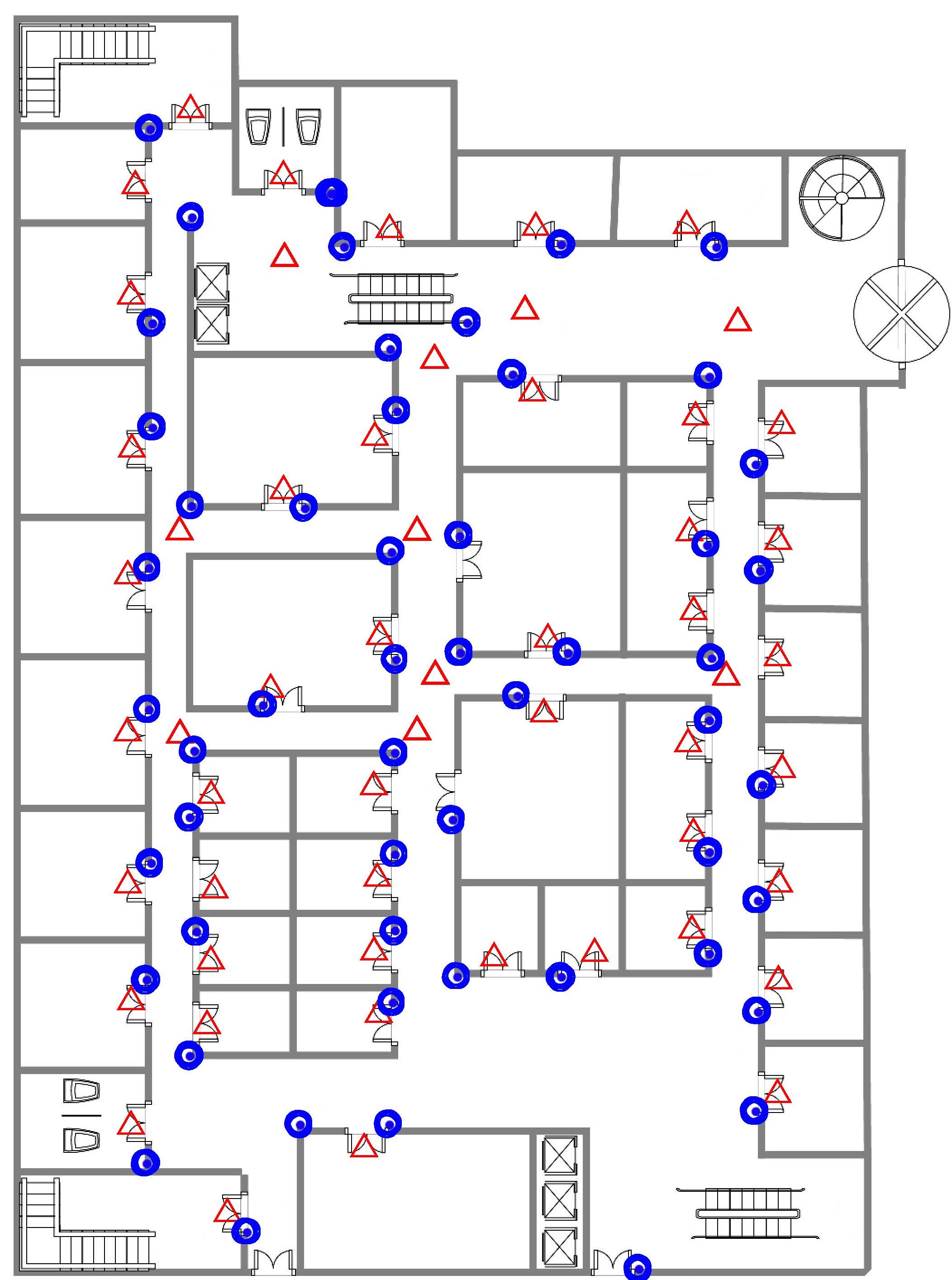}
		\subcaption{Beacon location markings on the Shopping Mall floor plan.}
		\label{fig:fp1marked}
	\end{minipage} \quad
	\begin{minipage}{0.49\textwidth}
		\centering
		\includegraphics[width=0.5\textwidth, height = 3.5in, angle = 90]{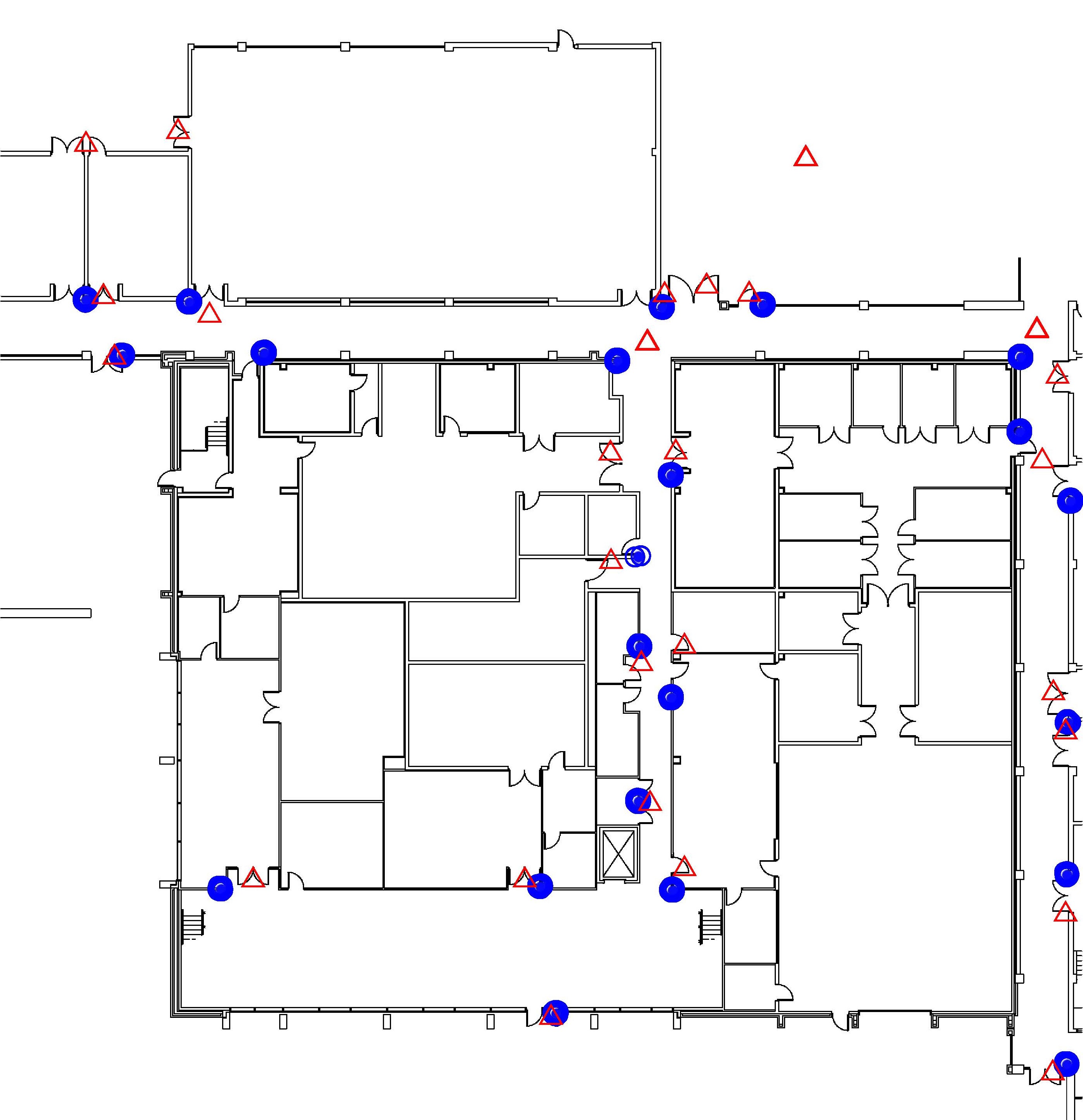}
		\subcaption{Beacon location markings on the Research Building floor plan.}
		\label{fig:fp2marked}
	\end{minipage}\quad
	\caption{Beacon locations as output on two different floor plans provided as input. Blue solid circles indicate beacon locations marked by a manual process while red triangles show beacon locations from IBeaconMap.}
	\label{fig:vaccuracy}
\end{figure*}

\subsection{Basic Results}
The basic results are from the FDM scheme, which is typically the recommended scheme (due to its low processing time) unless the floor plans have low resolution or very high density of building blocks. The basic results presented here use the indoor walking path detection of building blocks only option as this option is expected to be more commonly used. 

Figure \ref{fig:vaccuracy} shows the beacon location marking results using FDM on two floor plans (more floor plans are considered in the next subsection) that were of high resolution; one from a shopping mall and the other a small research building with offices and laboratories. For each floor plan, beacon locations identified by a manual process (finding all PoIs visually and marked) are shown along with those generated by IBeaconMap. It can be seen that the outputs are remarkably accurate. The major difference can be seen as the small mismatch in locations at each point of interest such as doors, stairs etc. which accounts for over half of an deviations seen and is easily correctable. This mismatch was because the manual process intentionally marks beacon locations on the side of a door or stairs while IBeaconMap marks them at the center leaving those who deploy the beacons to make the decision as to which side to place the beacon. In addition IBeaconMap marks additional locations at intersections which would be very useful during navigation. If two PoIs are very close to each other (within 2 m), IBeaconMap just affixes one beacon location that can serve both points. Some PoIs that perhaps would have been omitted as beacon locations during a manual process (due to knowledge that those PoIs will not be useful), are marked by IBeaconMap in the shown image; such location can be removed manually as explained in the next section.   

The computation time for both indoor floor plans considered in Figure \ref{fig:vaccuracy} were analyzed on an Intel i5-5200U CPU (2.20GHz) with
8 GB RAM on a 64-bit Windows 10 OS. The Shopping Mall floor plan took 15.64 seconds in total to provide the final outcome while Research Building only took 22.07 seconds. The computation for the former is faster as it has fewer building blocks/features that need to be detected next to the indoor path. Thus, these typical floor plans can be analysed to not only provide beacon locations to use, but they also provide a connectivity graph for navigation in under 1 minute. An entire building with multiple floors thus can be analyzed (and generation of connectivity information and beacon locations) in an automated fashion in the order of minutes to a few hours depending on its size. A manual process, that involved drawing walking paths, marking beacon locations, measuring and entering graph data structure connectivity information, weights, and directional orientations as experienced by the authors for the research building floor plan in \cite{GuideBeacon-percom2017}, took over 1 hour to arrive at similar outcomes; larger buildings with many more PoIs would have taken many more hours if not days per floor. It is important to remember that beacon planning is not just about sticking the beacons at some reasonable locations and using it for navigation; it requires a fair amount of computation on the back end to decide what locations will best serve navigational needs. Also, there may be post-deployment alterations required for which an automated tool again can make changes easier as explained in detail in the following section.

\begin{table*}[ht]
	\centering
	\caption{Comparison of Building Block Detection Techniques - Indoor Path Only}
	\label{tb:indoorplan}
	\resizebox{\textwidth}{!}{%
		
		\begin{tabular}{|l|l|c|c|c|c|l|l|c|c|c|c|c|}
			\hline
			\multicolumn{2}{|c|}{\multirow{2}{*}{}} & \multicolumn{1}{c|}{\multirow{2}{*}{Correct}} & \multicolumn{2}{c|}{Incorrect} & \multirow{2}{*}{Processing Time (s)} & \multicolumn{2}{c|}{\multirow{2}{*}{}} & \multirow{2}{*}{Correct} & \multicolumn{2}{c|}{Incorrect} & \multicolumn{2}{c|}{\multirow{2}{*}{Processing Time (s)}} \\ \cline{4-5} \cline{10-11}
			\multicolumn{2}{|c|}{} & \multicolumn{1}{c|}{} & Missed & Redundant &  & \multicolumn{2}{c|}{} &  & Missed & Redundant & \multicolumn{2}{c|}{} \\ \hline
			\multirow{3}{*}{Research Building} & FDM & 30 & 1 & 3 & 22.07 & \multirow{3}{*}{Large Area} & FDM & 105 & 1 & 23 & \multicolumn{2}{c|}{173.62} \\ \cline{8-8}
			& FDM+SML & 29 & 2 & 2 & 40.5 &  & FDM+SML & 98 & 8 & 16 & \multicolumn{2}{c|}{282.83} \\ \cline{8-8}
			& FD+SML & 31 & 0 & 12 & 51.94 &  & FD+SML & 105 & 1 & 18 & \multicolumn{2}{c|}{285.79} \\ \hline
			\multirow{3}{*}{Shopping Mall} & FDM & 49 & 0 & 12 & 15.64 & \multirow{3}{*}{Low-Resolution Image} & FDM & 22 & 9 & 6 & \multicolumn{2}{c|}{14.46} \\ \cline{8-8}
			& FDM+SML & 49 & 0 & 4 & 41.46 &  & FDM+SML & 21 & 10 & 3 & \multicolumn{2}{c|}{26.1} \\ \cline{8-8}
			& FD+SML & 49 & 0 & 1 & 44.41 &  & FD+SML & 31 & 0 & 15 & \multicolumn{2}{c|}{38.80} \\ \hline
			
	\end{tabular}}
\end{table*}

\begin{table*}[ht]
	\centering
	\caption{Comparison of Building Block Detection Techniques - Full Floor Plan}
	\label{tb:fullplan}
	\resizebox{\textwidth}{!}{%
		
		\begin{tabular}{|l|l|c|c|c|c|l|l|c|c|c|c|c|}
			\hline
			\multicolumn{2}{|c|}{\multirow{2}{*}{}} & \multicolumn{1}{c|}{\multirow{2}{*}{Correct}} & \multicolumn{2}{c|}{Incorrect} & \multirow{2}{*}{ Processing Time} & \multicolumn{2}{c|}{\multirow{2}{*}{}} & \multirow{2}{*}{Correct} & \multicolumn{2}{c|}{Incorrect} & \multicolumn{2}{c|}{\multirow{2}{*}{Processing Time}} \\ \cline{4-5} \cline{10-11}
			\multicolumn{2}{|c|}{} & \multicolumn{1}{c|}{} & Missed & Redundant &  & \multicolumn{2}{c|}{} &  & Missed & Redundant & \multicolumn{2}{c|}{} \\ \hline
			\multirow{3}{*}{Research Building} & FDM & 63 & 4 & 16 & 11.30 & \multirow{3}{*}{Large Area} & FDM & 237 & 11 & 97 & \multicolumn{2}{c|}{185.73} \\ \cline{8-8}
			& FDM+SML & 60 & 7 & 14 & 31.87 &  & FDM+SML & 233 & 15 & 59 & \multicolumn{2}{c|}{351.38} \\ \cline{8-8}
			& FD+SML & 67 & 0 & 9 & 67.94 &  & FD+SML & 248 & 0 & 68 & \multicolumn{2}{c|}{875.97} \\ \hline
			\multirow{3}{*}{Shopping Mall} & FDM & 49 & 0 & 40 & 18.77 & \multirow{3}{*}{Low-Resolution Image} & FDM & 42 & 25 & 30 & \multicolumn{2}{c|}{6.6} \\ \cline{8-8}
			& FDM+SML & 49 & 0 & 27 & 25.18 &  & FDM+SML & 42 & 25 & 20 & \multicolumn{2}{c|}{21.3} \\ \cline{8-8}
			& FD+SML & 49 & 0 & 0 & 81.33 &  & FD+SML & 67 & 0 & 13 & \multicolumn{2}{c|}{45.99} \\ \hline
			
	\end{tabular}}
\end{table*}

\subsection{Comparison of Building Block Detection Techniques}
All the building block detection techniques: FDM, FDM + SML, and FD + SML were compared in terms of statistical identification accuracy and processing time. These comparisons were done for the two cases described earlier in Section \ref{subsubsec:phase1}: detection along indoor path only, and entire floor plan detection. 

Evaluation results from four different floor plans are shown next. Many other floor plans were analyzed and tested to ensure that the results shown here are representative of a larger trend. The first two (Research Building and Shopping Mall) are those already seen in Figures \ref{fig:fp1marked} and \ref{fig:fp2marked}. An additional two, called Large Area and Scanned Image respectively, were added. The Large Area floor plan is of a 75,000 sq. ft indoor facility with a large number of potential PoIs, some of which are densely congregated as well. The fourth floor plan was the same as the first (Research Building), but a low-resolution (200 dpi) scanned image. These Large Area and Scanned Image floor plans were used to test the worst case for FDM and see how the SML based algorithms helped in such cases.

\subsubsection{Indoor path detection only}
The results for the detection along indoor walking path only is shown in Table \ref{tb:indoorplan}. It can be seen that all three building block detection schemes perform with a high accuracy in terms of correctly identifying PoIs with very few missed detections. The fast FDM scheme does very well for the smaller and simpler floor plans (Research Building and Small Area) and looks adequate for such cases. The FD + SML scheme helps improve detection accuracy significantly in the case of the low-resolution scanned image where FDM does not do well. The FD + SML scheme also seems to work better than FDM for floor plans with high density as in the Large Area floor plan. The FDM + SML scheme acts primarily as an ``enhancer'' to the FDM scheme, helping reduce some of the redundant locations identified, sometimes however at the cost of adding some more to missed detections. All schemes have some redundant identifications (false positives) which will need to be ``scrubbed off'' through a post-processing step as outlined in the following section. In terms of processing time (run on the same machine listed in the previous section), all techniques took only an order of seconds to minutes, with FDM being the fastest and FD + SML typically taking the most time. 

\subsubsection{Full floor plan detection}
The results for the entire floor plan building block detection is shown in Table \ref{tb:fullplan}. This being the worst case for building block detection due to the presence of multiple layers, it can be seen that the number of redundant beacon locations identified are larger; however, most PoIs are still correctly identified. The FD + SML scheme again improves upon that of the FDM scheme when image resolution is poor or has a high density of PoIs. The relative processing times of each scheme remains the same as in the indoor path only case, except that there is an overall increase due to the consideration of the entire floor area. As the floor plan area increases (as in the Larger Area floor plan), the FD + SML scheme processing time does increase faster than the other schemes due to its need to execute its three step process.

\begin{figure*}
	%\centering
%	\begin{minipage}{0.32\textwidth}
%		\centering
%		\includegraphics[width=\textwidth, height = 2in]{restAreaTool.jpg}
%		\subcaption{Option to mark any restricted access areas.}
%		\label{fig:RestArea}
%	\end{minipage} \quad
	\begin{minipage}{0.49\textwidth}
		\centering
		\includegraphics[width=\textwidth, height = 2in]{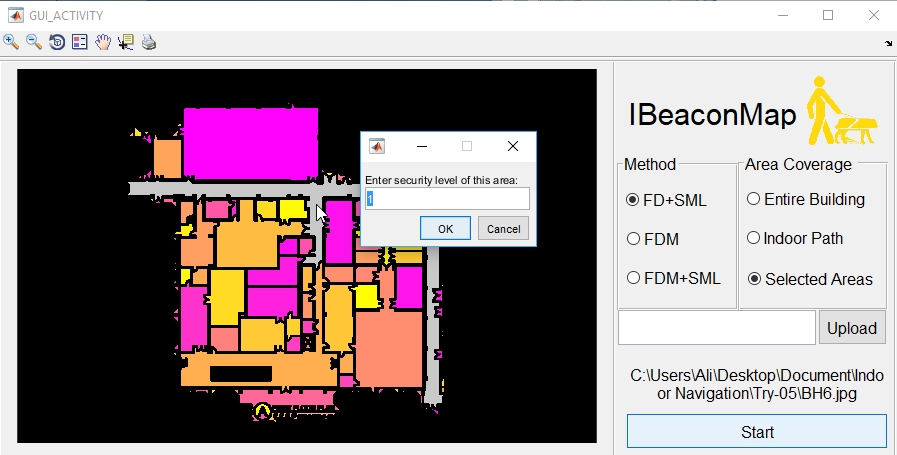}
		\subcaption{Marking restricted areas and assigning restriction levels.}
		\label{fig:selectingOption1}
	\end{minipage}\quad
\begin{minipage}{0.49\textwidth}
	\centering
	\includegraphics[width=\textwidth, height = 2in]{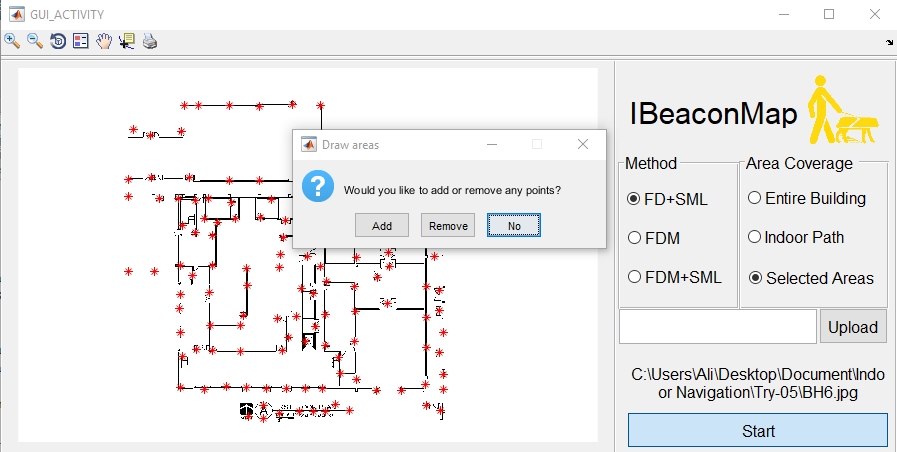}
	\subcaption{Post-processing option to correct any errors on beacon markings on floor plan.}
	\label{fig:tool}
	\end{minipage}
	\caption{Snapshots of IBeaconTool options.}
	\label{fig:toolimages}
\end{figure*}

\section{IBeaconMap Tool}\label{sec:tool}
This section describes the capabilities of the IBeaconMap software tool and how it can be used to carry out beacon planning of a site. 

\begin{figure}[htbp]
	{
		\includegraphics[width=\linewidth,height = 2in]{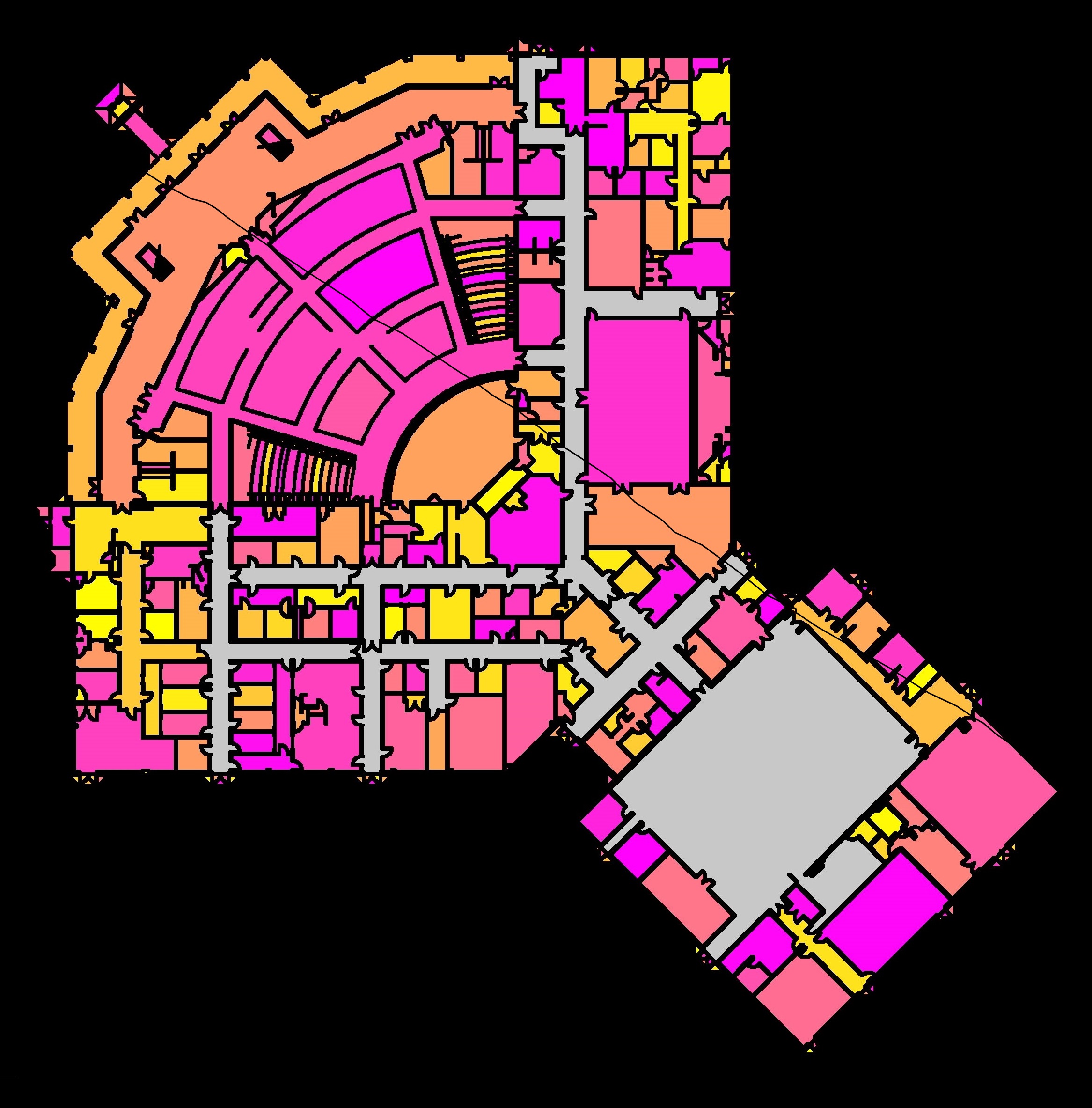}
		\caption{Zones within the floor plan of a large single story 75,000 sq. ft indoor facility set to various restriction levels.}
		\label{fig:metroplex}
	}
\end{figure}

The IBeaconMap tool provides various options for a user to pre-process floor plans that are being input, if they choose to do so. Common pre-processing tasks could include: (i) cropping the image to remove annotations and other symbols outside the floor plan boundaries, (2) marking certain areas to be protected\footnote{These could be restricted areas where a typical person is not allowed to enter such as a danger zone, or high-security area.}  and outside the bounds of navigation, and (3) marking off areas that should not be considered as part of the walking paths due to furniture or other objects. Levels of restricted areas are defined such that there are shades between full public access and no public access. For example, within an airport, the secure zones have restrictions for entry from many doors, but do allow walking once entered through an identified entryway. Figure \ref{fig:metroplex} shows the floor plan of a large metroplex with such restricted areas defined.

Once a floor plan image is uploaded, the user is provided three options for building block detection. The FDM method is recommended for cases where the image resolution is high enough and floor plan is of simple and small space. If these criteria are not met, the FD + SML method is recommended, though it can take extra time for processing. The FDM + SML option is a good alternative to FDM, although it takes a bit more time to process. A user can run all three options and compare if they choose to do so. In addition, the user is asked if they want just indoor walking path detection or a full floor plan detection of PoIs. If there are secure/restricted areas to be marked, that is done first before proceeding with the indoor path or full floor plan detection.

As post-processing steps, the IBeaconMap tool allows (i) "scrubbing off" any redundant locations that should not be beacon locations, while allowing adding of new locations which were possibly missed during processing, just by clicking with a mouse pointer, and (ii) labeling beacon locations with the descriptive detail and context required for navigating. 

Figure \ref{fig:toolimages} shows two snapshots of the various options provided within the tool. A demonstration video of how the IBeaconMap tool works is available to see at the site \url{http://wsumapping.cs.wichita.edu/IBeaconMap}.

\section{Discussion and Future Work}
%benefits

This work presented a largely automated technique called IBeaconMap to prepare an indoor space for beacon-based wayfinding for the BVI and other sighted users. Such a technique solves the current challenge of creating indoor space representations in a time and labor-efficient manner. IBeaconMap simply takes a floor plan of the indoor space under consideration and employs a combination of computer vision and machine learning techniques to arrive at locations where beacons can be deployed and builds a weighted connectivity graph among these locations that can be subsequently used for navigation. Evaluations show IBeaconMap to be fast computationally and reasonably accurate (depending on input resolution and space characteristics) thus presenting itself as a scalable tool in preparing all indoor spaces for beacon-based wayfinding in the future.  The actual placement of beacons can then be done very easily, not even requiring skilled technicians. Even crowd-sourcing approaches such as used in \cite{gleason2017luzdeploy} can be used to arrive at the final infrastructure required for beacon-based wayfinding. The merits of IBeaconMap lies not just in how it makes beacon planning easy for new deployments, but also the maintenance features it provides in making future changes and alterations. Its ability to handle low resolution images will allows it to be applied to scanned images of floor plans from older buildings and even hand-drawn sketches that follow similar standards.

%\newpage
\bibliographystyle{abbrv}
\bibliography{bibIndoorWayfindingAssets17-acm}  % sigproc.bib is the name of the Bibliography in this case
% ACM needs 'a single self-contained file'!
%\balancecolumns
%\appendix
%Appendix A

\end{document}